\documentclass[letterpaper]{article} 
\usepackage{aaai23}  
\usepackage{times}  
\usepackage{helvet}  
\usepackage{courier}  
\usepackage[hyphens]{url}  
\usepackage{graphicx} 
\usepackage{natbib}  
\usepackage{caption} 
\usepackage{algorithm}
\usepackage{algorithmic}

\usepackage{newfloat}
\usepackage{listings}
\DeclareCaptionStyle{ruled}{labelfont=normalfont,labelsep=colon,strut=off} 
\floatstyle{ruled}
\newfloat{listing}{tb}{lst}{}
\floatname{listing}{Listing}
\title{CKS: A Community-based K-shell Decomposition Approach using Community Bridge Nodes for Influence Maximization (Student Abstract)}
\author{
    Inder Khatri\equalcontrib \textsuperscript{\rm 1}, Aaryan Gupta\equalcontrib \textsuperscript{\rm 2}, Arjun Choudhry\equalcontrib \textsuperscript{\rm 1}, Aryan Tyagi\equalcontrib \textsuperscript{\rm 2}, Dinesh Kumar Vishwakarma\textsuperscript{\rm 1}, Mukesh Prasad\textsuperscript{\rm 3}
}
\affiliations{
    \textsuperscript{\rm 1} Biometric Research Laboratory, Delhi Technological University, New Delhi, India\\
    \textsuperscript{\rm 2} Delhi Technological University, New Delhi, India\\
    \textsuperscript{\rm 3} School of Computer Science, FEIT, University of Technology Sydney, Sydney, Australia\\
    \{inderkhatri999, aryan227227, choudhry.arjun, tyagiaryan82\}@gmail.com, dinesh@dtu.ac.in, mukesh.prasad@uts.edu.au
}

\usepackage{bibentry}

\begin{document}

\maketitle

\begin{abstract}
Social networks have enabled user-specific advertisements and recommendations on their platforms, which puts a significant focus on Influence Maximisation (IM) for target advertising and related tasks. The aim is to identify nodes in the network which can maximize the spread of information through a diffusion cascade. We propose a community structures-based approach that employs K-Shell algorithm with community structures to generate a score for the connections between seed nodes and communities. Further, our approach employs entropy within communities to ensure the proper spread of information within the communities. We validate our approach on four publicly available networks and show its superiority to four state-of-the-art approaches while still being relatively efficient.
\end{abstract}

\section{Introduction}
Online Social Networks are platforms for people to publicize their ideas and products. This approach, called Viral Marketing, raises the problem of Influence Maximisation (IM), which requires selecting $k$ seed nodes to maximize the information spread in the network. Several approaches have been proposed for IM in recent years, based on local, semi-local, and global structures. Global structure-based approaches are efficient due to their consideration of whole network. They find out core nodes with maximum connectivity to the remaining network. However, in real-world situations, where number of seed nodes is very small, influence gets restricted to only a few sub-groups (or communities) in the network containing the selected core seed nodes. To overcome this, we instead consider community bridge nodes as influential seed nodes due to their connections to a larger number of communities, leading to simultaneous information propagation to these communities. We propose CKS centrality measure, which incorporates community structures and K-shell Decomposition to identify influential spreaders in a network. We define three novel measures: Community K-Shells, Community K-Shell Entropy, and CKS-Score, which qualitatively and quantitatively evaluate connections of a node to various communities.

\section{Proposed Methodology}

\textbf{Obtaining Community K-Shell (CKS):}
Community K-Shell concept incorporates knowledge of information flow in a network. Obtaining CKSs comprises the following steps: identifying community structures using Louvain’s algorithm \citep{louvain}; isolating communities by removing connections between different communities; and passing isolated communities through K-Shell algorithm \citep{kshell} to obtain K-Shell scores particular to the community of each node (i.e., Community K-shell score). The higher the Community K-shell score of a node, the closer it is to the core of the given community.

\textbf{Computing K-Shell Entropy (KSE):}
KSE evaluates connectivity of a node to different regions of a community. We formulate KSE for a node $v$ corresponding to each community $c$ as follows: 
\begin{equation}
\label{eq:shells}
    \scriptstyle KSE_{v,c} \ = \ - \ \sum_{s = 1}^{shells_{c}} \ K_{s} \ * \ \frac{\eta_{v,s}}{\eta_{v}} \ * \ log(\frac{\eta_{v,s}}{\eta_{v}})
\end{equation}
\begin{equation}
\label{eq:eta}
    \scriptstyle \eta_{v} = \sum_{s'=1}^{shells_{c}}\eta_{v,s'}
\end{equation}
where, $\eta_{v,s}$ is number of connections of node $v$ with shell $s$ for a given community, $shells_{c}$ represents the unique CKS in community $c$, and $K_{s}$ represents K\_value of given shell.

Influence of an activated node lasts only till 2-3 hops. Thus, for maximum influence over a given community, connectivity of a node to the community should be well-distributed across all its regions, which also reduces the chances of overlapping influence in a community. This requires a higher KSE score.
Equation \ref{eq:shells} represents the entropy submission over shells of the respective community weighted by the K-Value of the respective shell.\\

\begin{figure*}[t!]
     \centering
     \includegraphics[width = 0.9\textwidth]{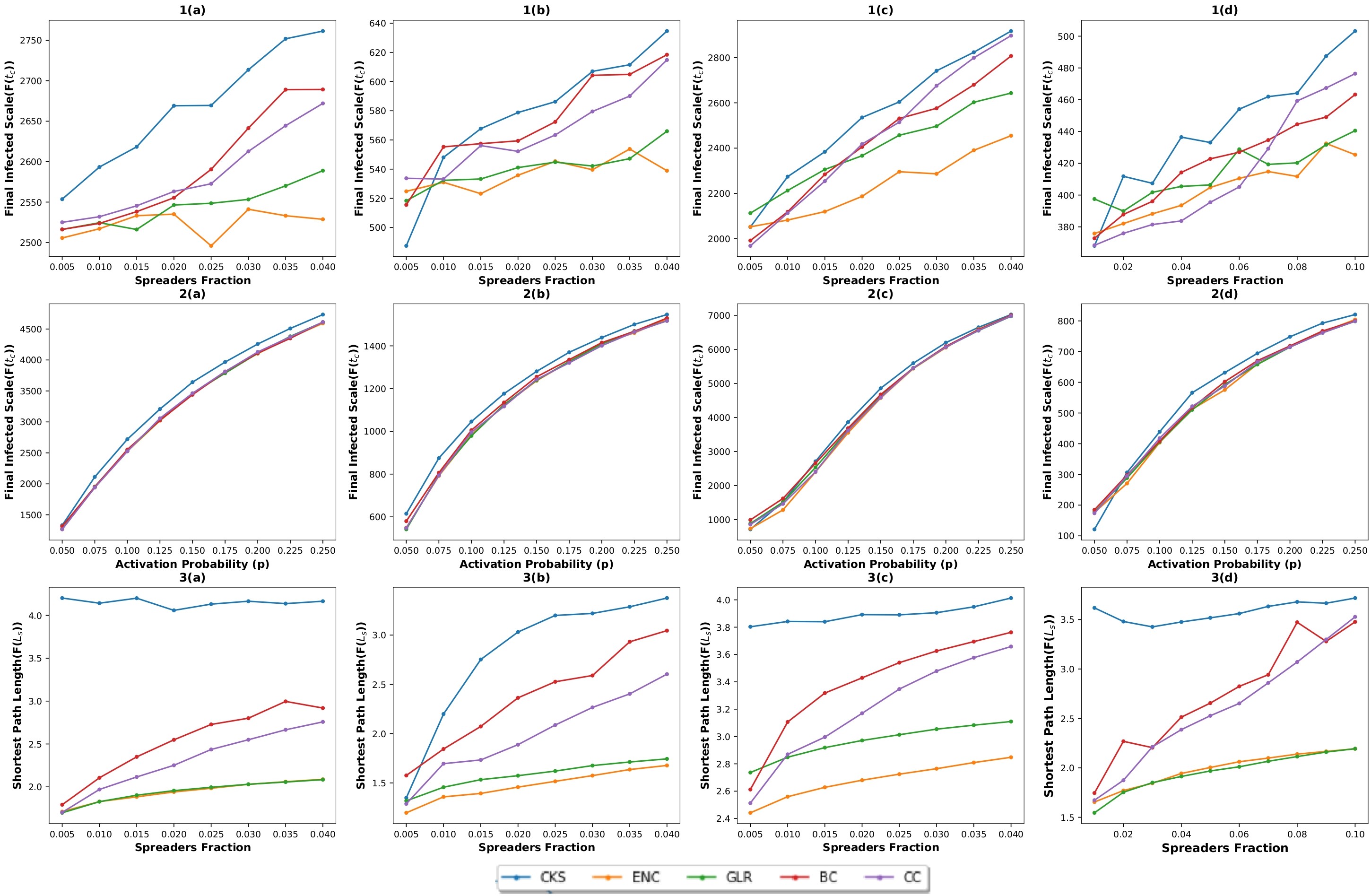}
     \hfill
   \caption{Results of our experiments on datasets (a) Twitch, (b) soc-Hamsterster (c) p2p-Gnutella04 (d) email-univ. (1) FIS after IC model simulation with activation probability 0.5. (2) FIS-v-p after IC model simulation with initial spreaders fraction 0.2. (3) ASPL among initial selected seed nodes vs spreaders fraction.}
   \label{Results} 
\end{figure*}

\textbf{CKS-Score:}
CKS-Score evaluates the overall connectivity of a node to all adjacent communities to which it is connected. It is defined as:
\begin{equation}
    \label{eq:comm}
    \scriptstyle CKS-Score(v) \ = \ \sum_{c = 1}^{comm} NN_{c} \ * \ KSE_{v,c} \ * \ \eta_{v}
\end{equation}
where, $\eta_{v}$ is number of connections of node $v$ with the respective community as shown in Equation \ref{eq:eta}, $KSE_{v,c}$ represents K-Shell Entropy for node $v$ and community $c$, and $NN_{c}$ represents number of nodes in community $c$.
We compute CKS-Score for each node by considering its KSE values corresponding to each community weighted by the community's size. This ensures a higher score for community bridge nodes connected to the most significant communities.

\section{Experimental Results and Discussion}
We evaluated CKS on four metrics: Final Infected Scale (FIS), Average Shortest Path Length (ASPL), Final infected scale vs Activation Probability (FIS-v-p), and Execution Time. We compared CKS with BC \citep{BC}, CC \citep{CC}, ENC \citep{ENC}, and GLR \citep{GLR} on four real-world datasets\footnotemark[1]: Twitch, soc-Hamsterster, p2p-Gnutella04, and email-univ. We simulated each model 100 times using Independent Cascade \citep{IC}. Infection probability was set to 0.1. Table \ref{ET} and Figure \ref{Results} show our findings. We observed:

\footnotetext[1]{https://networkrepository.com , https://snap.stanford.edu/snap}

\begin{table}[t!]
    \centering
    \large
    \resizebox{\columnwidth}{!}{
    \begin{tabular}{c|ccccc}
    \hline
    \hline
        \textbf{Dataset} & \textbf{CKS} & \textbf{ENC} & \textbf{GLR} & \textbf{BC} & \textbf{CC} \\\hline
        Twitch & 28.97 & 0.176 & 37.72 & 533.63 & 159.07 \\
        soc-Hamsterster & 49.21 & 0.04 & 3.44 & 40.12 & 11.84 \\
        p2p-Gnutella04 & 83.21 & 0.19 & 54.41 & 1205.83 & 352.65 \\
        email-univ & 0.83 & 0.01 & 0.71 & 10.09 & 1593.18 \\
    \hline
    \hline
    \end{tabular}}
    \caption{Execution time for CKS and other approaches.}
    \label{ET}
\end{table}

CKS consistently outperformed other core nodes-based approaches on FIS, ASPL, and FIS-v-p, while being reasonably efficient. This confirms our hypothesis that bridge nodes are more influential core nodes.

CKS showed a higher inter-spreader distance than competing approaches, while also having a higher infection rate, leading to much lesser overlap between the influence of spreaders (or \emph{rich club effect}).

\nocite{*}
\fontsize{9pt}{10pt}\selectfont {\bibliography{main.bib}}

\end{document}